\documentclass[11pt]{article}
\usepackage{amssymb,amsmath,amsfonts}
\usepackage{graphicx}
\usepackage{graphics}
\usepackage{eepic,epsfig}

1.60cm \makeatletter \@addtoreset{equation}{section}

\makeatother
\begin{document}

\title{Azimuthal fermionic current in the cosmic string spacetime induced by a magnetic tube}
\author{M. S. Maior de Sousa \thanks{E-mail: kaelsousa@gmail.com}\ ,  R. F. Ribeiro \thanks{E-mail: 
rfreire@fisica.ufpb.br} and  E. R. Bezerra de Mello \thanks{E-mail: emello@fisica.ufpb.br} \
\\
Departamento de F\'{\i}sica-CCEN\\
Universidade Federal da Para\'{\i}ba\\
58.059-970, J. Pessoa, PB\\
C. Postal 5.008\\
Brazil}

\maketitle

\begin{abstract}
In this paper, we analyze the vacuum azimuthal fermionic current induced by a 
magnetic field confined in a cylindrical tube of finite radius $a$, in the 
cosmic string spacetime. Three distinct configurations for the magnetic field are taken into account: (i) a cylindrical  shell of radius $a$, (ii) a magnetic field proportional to $1/r$ and (iii) a constant 
magnetic field. In these three cases, the axis of the infinitely long tube of radius $a$ coincides with the cosmic 
string; moreover, we only develop this analysis for the region outside the tube. 
In order to do that, we explicitly construct the corresponding complete set of normalized wave-functions. 
We show that in the region outside the tube, the induced current is
decomposed into a part corresponding to a zero-thickness magnetic flux in addition 
to a core-induced contribution. The latter presents specific form depending on the 
magnetic field configuration considered. The zero-thickness contribution depends only on the fractional part of 
the ration of the magnetic  flux inside the tube by the quantum one. As to the 
core-induced contribution, it depends on the  total magnetic flux inside the tube, and consequently, in general, 
it is not a periodic function of the flux. 
\end{abstract}

\bigskip


\bigskip

\section{Introduction}
\label{Int}

The existence of a magnetic flux tube penetrating a type II superconductor,
named {\bf vortex}, was first demonstrated by Abrikosov
\cite{Abrikosov}, by using the Ginzburg-Landau theory of superconductivity.
Some years later, Nielsen and Olesen \cite{nielsen} have shown, 
by using a classical relativistic field theory, composed by Higgs fields
interacting with Abelian one, that presents
spontaneously gauge symmetry broken, contains static cylindrically symmetric
solution carrying a magnetic flux. This configuration corresponds to the vortex solution.
The equations of motion associated for this system form a set of coupled
non-linear differential equation, that, in general, has no closed solutions. The analysis
of the influence of this system on the geometry of the spacetime was analyzed
numerically by Garfinkle  \cite{DG} and Laguna \cite{Laguna} many years ago. In
these papers the authors showed that, the vortex has a inner structure
characterized by a non-vanishing core carrying a magnetic flux, whose
extension is determined by the energy scale where the symmetry is broken. Two
length scales naturally appear, the one related with the extension of the
magnetic flux proportional to the inverse of vector field mass, $m_v$, and
the other associated with the inverse of the scalar field mass, $m_s$. The latter
being the radius that the scalar field reaches its vacuum
value. Moreover the authors also verify that asymptotically the surface
perpendicular to the vortex corresponds to a Minkowski one minus a wedge.

According to the Big Bang theory, during its expansion the universe underwent to a 
series of phase transition. In most interesting model of high-energy physics, the formation
of topological defects such as domain  walls, monopoles, cosmic string, among others 
are predicted  to occur \cite{vilenkin}. The cosmic string, a linear topological defect, 
is the most studied. Though the recent observations of the cosmic microwave background radiation have 
ruled out them as the primary source for primordial density perturbations, cosmic 
strings give rise to a number of interesting physical effects such as gamma rays bursts \cite{Bere}, the emission of 
gravitational waves \cite{Damour} and the generation of high-energy cosmic rays \cite{sigl}. 
String-like defects also appear in a number of condensed matter systems, including liquid crystals 
and graphene made structures.

The complete analysis of the behavior of a quantum charged field in the neighborhood  a 
Nielsen and Olesen (NO) string must take into account the influence of the geometry of the
spacetime and also the presence of the magnetic field. Two distinct approach for 
this system are: $(i)$ To consider the string as an idealized linear topological defect,
having a magnetic field running along it. This case can be treated analytically. 
$(ii)$ To consider the non-zero thickness for the string. Unfortunately this problem 
is analytically intractable. In the idealized model, the conical structure of the spacetime
modifies the zero-point vacuum fluctuation of quantized fields inducing non-vanishing vacuum
expectation values (VEV) for important physical observable, such as the energy-momentum tensor. 
See for instance references given in \cite{Beze06sc}, for the cases of uncharged scalar, 
fermionic and vector fields. Considering the presence magnetic flux, additional polarization effects 
associated with charged quantum fields take place \cite{Dowk87}-\cite{Site12}. In particular 
this magnetic flux induces non-vanishing current densities, $\langle j^\mu\rangle$. This phenomenon 
has been investigated for scalar fields in Ref.\cite{Srir01,Site09}. The analysis of 
induced fermionic currents in higher-dimensional cosmic string spacetime in the presence of a magnetic flux have 
been developed in Ref. \cite{Mello10a}. In all these analysis, the authors have shown 
that induced azimuthal vacuum current densities take place if the ratio of the magnetic flux by the quantum one 
has a nonzero fractional part.\footnote{Moreover, induced fermionic and scalar current 
	densities in compactified cosmic string spacetimes have been considered in \cite{Saha-Mello} and \cite{Brag}.}

Because the analysis of a quantum system in a realistic model for the NO vortex
cannot be exactly solvable, an intermediate approach can be adopted. Here we assume an  
approximated model that consists to consider the spacetime produced by the string as 
being conical everywhere, but having a non-zero thickness magnetic field surrounding it. In this way,
some improvements can be obtained. This approach was used in \cite{Spin03,Spin04} to 
calculate the VEV of massless charged scalar and fermionic energy-momentum tensors, 
$\langle T_{\mu\nu}\rangle$, respectively. More recently the scalar vacuum current induced by a magnetic flux 
in a cosmic string considering a non-vanishing core has been developed in \cite{Mello15}. 
In these calculations the vacuum polarization effects were developed for the region outside the tube, and shown to present two contributions. The first one associated with a 
zero-thickness magnetic flux, and the second induced by the non-vanishing core, named 
core-induced contributions. The first contribution is a periodic function  of the magnetic flux inside the tube, with the period equal to the quantum flux, $\Phi_0=2\pi/e$; as to the second, in general, are not periodic function 
of the quantum flux. In fact the core-induced contributions depends on the total magnetic flux inside the core. 
The later is  new  one, and its corresponding result may shed light upon features of finite core effects in 
more realistic models. Three different configurations of magnetic flux will be 
considered: $(i)$ A magnetic field on a cylindrical shell, $(ii)$ a magnetic field proportional to $1/r$, 
and finally $(iii)$ a homogeneous magnetic field inside the tube.
	
This paper is organized as follow. In section \ref{Geometry} we describe the background geometry 
of the spacetime and the configurations of the magnetic fields. We provide the general structure of 
the  complete set of normalized positive- and negative-energy fermionic mode functions in the region outside the tube, which can 
be used for each magnetic field configuration.  In section \ref{Mode}, by using the mode-summation 
method, we calculate the induced  azimuthal current density.  We show that this current can be 
decomposed in two parts: The first one corresponds to the induced current by a zero-thickness 
magnetic flux in the geometry of an idealized cosmic string, and the second one is induced by the 
non-zero core of the magnetic tube.
The latter is calculated, separately, for the three kinds of magnetic flux considered. 
The behavior of them are discussed in various asymptotic regions of the parameters. We also 
present some plots associated with  the core-induced azimuthal current exhibiting its behavior as function 
of the most relevant physical variables. Our most relevant conclusions are summarized in section \ref{Conclu}.

\section{The geometry and the fermionic wave-functions}
\label{Geometry}
The background geometry associated with an idealized cosmic string along the $z$-axis, can 
be given, by using cylindrical coordinates, through the line element below:
\begin{equation}
\label{2.1}
ds^2=dt^2 - dr^2 - r^2d\phi^2-dz^2 \ ,
\end{equation}
where the coordinates take values in range $r\geq 0$, $0\leq\phi\leq\phi_0=2\pi/q$ and 
$-\infty\leq (t, \ z)\leq+\infty$. The parameter $q$ associated with the planar angle 
deficit is related to the mass per unit length of the string, $\mu_0$,  by $q^{-1}=1-4\mu_0$.

The quantum dynamic of a massive charged spinor field in curved space-time and in the
presence of an electromagnetic four-vector potential, $A_\mu$, is described by the Dirac equation,
\begin{equation}
\label{2.2}
i\gamma^{\mu} (\nabla_\mu + ieA_\mu)\psi-m\psi=0, \ \ \nabla_\mu=\partial_\mu + \Gamma_\mu \ ,
\end{equation}
where $\gamma^\mu$ are the Dirac matrices in curved space-time and $\Gamma_\mu$ is the spin
connection. For the geometry in consideration the gamma matrices can be written in the form,
\begin{equation}
\label{2.6}
\gamma^0=\gamma^{(0)}=\left( \begin{array}{cc}
1 & 0 \\
0 & -1 \end{array} \right), \ \ \gamma^{l}=\left( \begin{array}{cc}
0 & \sigma^l \\
-\sigma^l & 0 \end{array} \right),
\end{equation}
where we have introduced the $2 \times 2$ matrices for $l=(r, \ \phi, \ z)$:
\begin{equation}
\label{2.7}
\sigma^{r}=\left( \begin{array}{cc}
0 & e^{-iq\phi} \\
e^{iq\phi} & 0 \end{array} \right), \ \sigma^{\phi}=-\frac{i}{r}\left( \begin{array}{cc}
0 & e^{-iq\phi} \\
-e^{iq\phi} & 0 \end{array} \right), \ \sigma^{z}=\left( \begin{array}{cc}
1 & 0 \\
0 & -1 \end{array} \right)  \  .
\end{equation}
As to the four vector potential we consider 
\begin{equation}
\label{2.01}
A_{\mu}=(0,0,A_\phi(r),0) \ ,
\end{equation}
with
\begin{equation}
\label{2.012}
A_\phi(r)=-\frac{q\Phi}{2\pi}a(r) \ .
\end{equation}
For the first model, representing a cylindrical magnetic shell,
\begin{equation}
\label{2.013}
a(r)=\Theta(r-a) \ .
\end{equation}
For the second and third models, representing a magnetic field proportional to $1/r$ and 
an homogeneous field, respectively, the radial function $a(r)$ reads,
\begin{equation}
\label{2.014}
a(r)=f(r)\Theta (a-r)+\Theta (r-a) \ ,
\end{equation}
with
\begin{eqnarray}
\label{2.015}
f(r)=\left\{\begin{array}{cc}
r/a,&\mbox{for the second model} \  ,  \\
r^2/a^2,&\mbox{for the third model} \  .
\end{array}
\right.
\end{eqnarray}
In the expressions above $\Theta(z)$ represents the Heaviside function, and $\Phi$ the total magnetic flux. 

For positive-energy solutions, assuming the time-dependence of the eigenfunctions in the form
$e^{-iEt}$ and decomposing the spinor $\psi$ into the upper $(\psi_{+})$ and lower $(\psi_{-})$
components, we find the following equations
\begin{equation}
\label{2.10}
(E-m)\psi_{+} + i\left[\sigma^l \left(\partial_l +ieA_l\right)+\frac{1-q}{2r}\sigma^r\right]\psi_{-}=0 \  ,
\end{equation}
\begin{equation}
\label{2.11}
(E+m)\psi_{-} + i\left[\sigma^l \left(\partial_l +ieA_l\right)+\frac{1-q}{2r}\sigma^r\right]\psi_{+}=0  \  .
\end{equation}
Substituting the function $\psi_{-}$ from the second equation into the first one, we
obtain the second order differential equation for the spinor $\psi_{+}$:
\begin{eqnarray}
\label{2.12}
\left[r^2\partial^2_r + r \partial_r + \left(\partial_\phi + ieA_\phi -
i\frac{1-q}{2}\sigma^z\right)^2+ \right.\nonumber\\ \left. r^2(\partial^2_z + E^2 - m^2)-
\frac{e}{r}\sigma^z \partial_r A_\phi\right]\psi_{+}&=&0  \   .
\end{eqnarray}
The same equation is obtained for $\psi_{-}$. So, we may say that the general solutions
to $\psi_+$ and $\psi_-$ can be express in terms of the ansatz below,
compatible with the cylindrical symmetry of the physical system,
\begin{equation}
\label{2.13}
\psi_{+}=e^{-ip.x}e^{iqn\phi}\left( \begin{array}{cc}
R_{1}(r) \\
R_{2}(r)e^{iq\phi} \end{array} \right) \  ,
\end{equation}
\begin{equation}
\label{1.14}
\psi_{-}=e^{-ip.x}e^{iqn\phi}\left( \begin{array}{cc}
R_{3}(r) \\
R_{4}(r)e^{iq\phi} \end{array} \right)\  ,
\end{equation}
where $p.x\equiv Et-kz$.
This function is eigenfunction of the total angular momentum
along the string, $J_z$,
\begin{equation}
\label{2.15}
\hat{J_z}\psi=\left(-i\partial_\phi+i\frac{q}{2}\gamma^{(1)}\gamma^{(2)}\right)\psi=qj\psi \ ,
\end{equation}
where $j=n+1/2, \ n=0, \ \pm 1, \ \pm 2, \ ...   \ .$

In order to construct the complete set of the wave-functions we shall consider, separately, 
the equation \eqref{2.12} in the regions $r<a$ and $r>a$. For the inner region, three different 
configurations of magnetic field have been already specified by the four-vector potential 
\eqref{2.01}-\eqref{2.015}. For the outer region, $r>a$, there is no magnetic field and 
the vector potential given by,
\begin{equation}
\label{A-B}
A_\phi=-\frac{q\Phi}{2\pi} \  ,
\end{equation}
being $\Phi$ the magnetic flux. So,  we find that the external positive-energy solution of the 
Dirac equation is given in terms of Bessel, $J_\mu(z)$, and Neumann, $Y_\mu(z)$, functions. A similar 
result is verified for the lower components of the Dirac spinor. Consequently we can write:
\begin{equation}
\label{2.19}
\psi^{(+)}=e^{-ip.x}e^{iqn\phi}\left( \begin{array}{c}
C_{1}J_{\beta_j}(\lambda r)+D_{1}Y_{\beta_j}(\lambda r) \\
\left[C_{2}J_{\beta_j+\epsilon_j}(\lambda r)+D_{2}Y_{\beta_j+\epsilon_j}(\lambda r)\right]e^{iq\phi} \\
A_{1}J_{\beta_j}(\lambda r)+B_{1}Y_{\beta_j}(\lambda r) \\
\left[A_{2}J_{\beta_j+\epsilon_j}(\lambda r)+B_{2}Y_{\beta_j+\epsilon_j}(\lambda r)\right]e^{iq\phi} \end{array} \right),
\end{equation}
with $n=j-1/2$ being an integer number. We have defined
\begin{equation}
\label{2.20}
\beta_j=q|j+\alpha|-\frac{\epsilon_j}{2} \   ,
\end{equation}
with $\epsilon_j=1$ for $j\geq-\alpha$ and $\epsilon_j=-1$ for $j<-\alpha$, being 
$\alpha=eA_\phi/q=-\Phi/\Phi_0$, Here $\Phi_0=2\pi/e$ is the quantum flux. As we can see, 
we have introduced a set of eight constants $C_i$, $D_i$, $A_i$ and
$B_i$ with $i=1, \ 2$ in the general solution above.

The energy is expressed in terms of $\lambda, \ k \ \mbox{and} \ m$ by the relation
\begin{equation}
\label{2.21}
E=\sqrt{\lambda^2 + k^2 + m^2}.
\end{equation}
We can find a relation between the constants of the the upper and lower solutions 
in (\ref{2.19}) by the use of (\ref{2.10}) and (\ref{2.11}). The relations are given by
\begin{equation}
\label{2.22}
A_1=\frac{kC_1-i\epsilon_j \lambda C_2}{E+m}, \ \ A_2=-\frac{kC_2-i\epsilon_j \lambda C_1}{E+m}
\end{equation}
\begin{equation}
\label{2.23}
B_1=\frac{kD_1-i\epsilon_j \lambda D_2}{E+m}, \ \ B_2=-\frac{kD_2-i\epsilon_j \lambda D_1}{E+m}.
\end{equation}
In addition, for the further specification of the eigenfunctions, we can impose extra conditions 
relating the above constants. As such a condition, following \cite{Nail}, we will
require the following relations between the upper and lower components:
\begin{equation}
\label{2.24}
R_3(r)=\rho_s R_1(r), \ \ R_4(r)=-\frac{R_2(r)}{\rho_s},
\end{equation}
with
\begin{equation}
\label{rho}
\rho_s=\frac{E+s\sqrt{\lambda^2 + m^2}}{k} \ , \  s= \pm 1  \ .
\end{equation}
Doing this we obtain the following relations:
\begin{eqnarray}
\label{rho1}
A_1=\rho_sC_1 \ , \ A_2=-C_2/\rho_s \  , \\
B_1=\rho_sD_1 \ ,  \  B_2=-D_2/\rho_s \  .
\end{eqnarray}
Hence, the positive frequency exterior solutions to the Dirac equation, specified by the 
set of quantum numbers $\sigma=(\lambda, \ j, \ k, \ s)$,  has the form
\begin{equation}
\label{out}
\psi^{(+)}_{\sigma(out)}(x)=e^{-ip.x}e^{iqn\phi}\left( \begin{array}{c}
C_{1}J_{\beta_j}(\lambda r)+D_{1}Y_{\beta_j}(\lambda r) \\
i\epsilon_j \rho_s b^{(+)}_{s}\left[C_{1}J_{\beta_j+\epsilon_j}(\lambda r)+
D_{1}Y_{\beta_j+\epsilon_j}(\lambda r)\right]  e^{iq\phi} \\
\rho_s\left[C_{1}J_{\beta_j}(\lambda r)+D_{1}Y_{\beta_j}(\lambda r)\right] \\
-i\epsilon_j b^{(+)}_{s}\left[C_{1}J_{\beta_j+\epsilon_j}(\lambda r)+
D_{1}Y_{\beta_j+\epsilon_j}(\lambda r)\right] e^{iq\phi} \end{array} \right),
\end{equation}
where we have introduced
\begin{equation}
\label{2.26}
b^{(\pm)}_s=\frac{\pm m+s\sqrt{\lambda^2+m^2}}{\lambda} \  .
\end{equation}

For the region inside, $r<a$, we have three different configurations of magnetic field,
as we have already mentioned. In this way we have three different solutions for
\eqref{2.12}. Let us represent each radial function by $R_l^{(i)}$, where $i=1, \ 2, \ 3$,
is the index associated with the model and $l=1, \ 2$  the index specifying the
the spinor components. For the inner region we can write the positive-energy
spinor field in the general form below:
\begin{equation}
\label{in}
\psi^{(+)}_{i(in)}(x)=C^{(i)} e^{-ip.x}e^{iqn\phi}\left( \begin{array}{c}
R^{(i)}_1(\lambda, r)\\
i\rho_s b^{(+)}_{s} R^{(i)}_2(\lambda, r) e^{iq\phi} \\
\rho_s R^{(i)}_1(\lambda, r) \\
-i b^{(+)}_{s}R^{(i)}_2(\lambda, r) e^{iq\phi} \end{array} \right) \   .
\end{equation}

The coefficients $C_1$ and $D_1$ in \eqref{out} and $C^{(i)}$ in \eqref{in} can be determined 
by the continuity condition of the fermionic wave function at $r=a$. After some intermediate
steps we can write,
\begin{eqnarray}
\label{Rela1}
C_1&=&-\frac\pi2(\lambda a)C^{(i)}R^{(i)}_1(\lambda, a){\tilde{Y}}_{\beta_j}(\lambda a) \ , \\
\label{Rela2}
D_1&=&\frac\pi2(\lambda a)C^{(i)}R^{(i)}_1(\lambda, a){\tilde{J}}_{\beta_j}(\lambda a) \  ,
\end{eqnarray}
where
\begin{eqnarray}
\label{Z.Bessel}
{\tilde{Z}}_{\beta_j}(z)=\epsilon_jZ_{\beta_j+\epsilon_j}(z)-{\cal{V}}_j^{(i)}(\lambda, a)
 Z_{\beta_j} (z)  \ , \  {\rm with} \
{\cal{V}}^{(i)}_j(\lambda, a)=\frac{R_2^{(i)}(\lambda, a)}{R^{(i)}_1(\lambda, a)} \ .
\end{eqnarray}
In \eqref{Z.Bessel} $Z_\mu$ represents the Bessel functions $J_\mu$ or $Y_\mu$.
With this notation all the informations about the inner structure of the magnetic field is contained in
the coefficient ${\cal{V}}_j^{(i)}$.

Finally the constant $C^{(i)}$ can be obtained form the normalization condition,
\begin{equation}
\label{Ren}
\int{d^3x\sqrt{g^{(3)}}\left(\psi^{(+)}_\sigma\right)^\dagger\psi^{(+)}_{\sigma'}}=
\delta_{\sigma,\sigma'} \ ,
\end{equation}
where delta symbol on the right-hand side is understood as the Dirac delta function for
continuous quantum numbers $\lambda \ \mbox{and} \ k$, and the Kronecker delta for discrete 
ones $n \ \mbox{and} \ s$, and $g^{(3)}$ is the determinant of the spatial metric tensor.
The integral over the radial coordinate should be done in the interval $[0, \ \infty)$. In this case 
two different expressions for the wavefunction must be used. The function  \eqref{in} for $r\in[0,a]$ 
and \eqref{out} for $r\in[0, \ \infty)$. The integral over the interior region is finite, consequently 
the dominant contribution for $\lambda'=\lambda$ comes from the integration in the exterior region. 
By using the standard integrals involving the cylindrical Bessel
functions, we find
\begin{equation}
\label{Rela3}
(2\pi)^2[|C_1|^2+|D_1|^2]=\frac{q\lambda}{(1+\rho^2_s)(1+({b^{(+)}_s})^2)}  \  .
\end{equation}
Substituting  \eqref{Rela1} and \eqref{Rela2} into the above equation we find:
\begin{eqnarray}
\label{Rela4}
C^{(i)}R^{(i)}_1(\lambda, a)=\Xi(\lambda,a) \  ,
\end{eqnarray}
with
\begin{eqnarray}
\Xi(\lambda,a)=\frac1{a\pi^2}\left[\frac q\lambda\frac1{(1+\rho^2_s)}
\frac1{(1+({b^{(+)}_s})^2)}\right]^{1/2} \frac1{\sqrt{(\tilde{Y}_{\beta_j}(\lambda a))^2
+(\tilde{J}_{\beta_j}(\lambda a))^2}} \  .
\end{eqnarray}
This relation determines the normalization constant for the interior wave function. 

The out-side negative-energy fermionic wave-function can be obtained in a similar procedure. 
So, the positive- and negative-energy wavefunctions are specified by the complete set of 
quantum numbers $\sigma=(\lambda, \ k, \ j, \ s)$. These functions can be written as show below:	
\begin{equation}
\label{psi-out}
\psi^{(\pm)}_{\sigma(out)}(x)=C^{(\pm)}_{(out)}e^{\mp i(Et-kz)}e^{iq(j-1/2) \phi}\left( \begin{array}{c}
g_{\beta_j}(\lambda a, \lambda r) \ ,  \\
\pm i\epsilon_j \rho_s b^{(\pm)}_s g_{\beta_j+\epsilon_j}(\lambda a, \lambda r) e^{iq\phi} \\
\rho_s g_{\beta_j}(\lambda a, \lambda r) \\
\mp i\epsilon_j b^{(\pm)}_s g_{\beta_j+\epsilon_j}(\lambda a, \lambda r) e^{iq\phi} \end{array} \right) \  ,
\end{equation}
where we have introduced the notations,
\begin{equation}
\label{g-beta}
g_{\beta_j}(\lambda a, \lambda r)=\frac{\tilde{Y}_{\beta_j}(\lambda a)J_{\beta_j}(\lambda r)
	-\tilde{J}_{\beta_j}(\lambda a)Y_{\beta_j}(\lambda r)}{\sqrt{(\tilde{Y}_{\beta_j}(\lambda a))^2
		+(\tilde{J}_{\beta_j}(\lambda a))^2}}  \  ,
\end{equation}
\begin{equation}
\label{g-beta1}
g_{\beta_j+\epsilon_j}(\lambda a, \lambda r)=\frac{\tilde{Y}_{\beta_j}(\lambda a)
	J_{\beta_j+\epsilon_j}(\lambda r)-\tilde{J}_{\beta_j}(\lambda a)Y_{\beta_j+\epsilon_j}
	(\lambda r)}{\sqrt{(\tilde{Y}_{\beta_j}(\lambda a))^2+(\tilde{J}_{\beta_j}(\lambda a))^2}}  \   ,
\end{equation}
\begin{eqnarray}
\label{rho}
\rho_s=\frac{E+s\sqrt{\lambda^2 + m^2}}{k} \ , \  s= \pm 1   \  , 
\end{eqnarray}
\begin{equation}
\label{2.26}
b^{(\pm)}_s=\frac{\pm m+s\sqrt{\lambda^2+m^2}}{\lambda} 
\end{equation}
and
\begin{equation}
\label{norm}
C^{(\pm)}_{(out)}=\frac{1}{2\pi}\left[\frac{q\lambda}{(1+\rho^2_s)\left(1+(b^{(\pm)}_s)^2\right)}\right]^{1/2}  \  ,
\end{equation}

Having the negative-energy wave-function for the region outside the magnetic flux, i.e.,
for $r>a$, we can use it for the investigation of the vacuum azimuthal  fermionic current density.

\section{Induced Azimuthal Fermionic current}
\label{Mode}
The vacuum expectation value (VEV) of the fermionic current density operator, 
$j^{\mu}=e\bar{\psi}\gamma^{\mu}\psi$, can be obtained by using the mode sum formula,
\begin{equation}
\label{current}
\left\langle j^\mu(x) \right\rangle=e\sum_{\sigma}{\bar{\psi}^{(-)}_\sigma (x)\gamma^\mu {\psi}^{(-)}_\sigma (x)}  \    ,
\end{equation}
where we are using the compact notation defined by
\begin{equation}
\label{sum}
\sum_{\sigma}=\int^{\infty}_{0} d\lambda \int^{+\infty}_{-\infty} dk \sum_{j=\pm 1/2, ...} \sum_{s=\pm 1}  \  .
\end{equation}

Specifically the VEV of the azimuthal current density is given by,
\begin{eqnarray}
\label{Azimuthal}
\langle j^\phi (x) \rangle=e\sum_{\sigma}{(\psi^{(-)}_\sigma (x))^\dagger
\gamma^0 \gamma^\phi {\psi}^{(-)}_\sigma (x)}  \  .
\end{eqnarray}
By using \eqref{psi-out}, \eqref{norm} and the explicit form of the Dirac
matrices given by (\ref{2.6}) and (\ref{2.7}) the following expression for the induced azimuthal current density in the region outside the tube is obtained:
\begin{eqnarray}
\label{Azimuthal2}
\langle j^\phi \rangle=-\frac{eq}{2\pi^2 r}\int^{\infty}_{-\infty} dk \int^{\infty}_{0}
\frac{\lambda^2d\lambda}{\sqrt{\lambda^2 + k^2 + m^2}}\sum_{j}\epsilon_j
g_{\beta_j}(\lambda a, \lambda r)g_{\beta_j+\epsilon_j}(\lambda a, \lambda r) \   .
\end{eqnarray}
Developing the product of $g_{\beta_j}(\lambda a, \lambda r)g_{\beta_j+\epsilon_j}(\lambda a, \lambda r)$
in a convenient form, i.e., separating the contributions that does not depend on the 
inner structure of the magnetic field from the other that does, we can written
the above result as the sum of two terms as shown below:
\begin{eqnarray}
\label{Azimuthal3}
\langle j^\phi (x) \rangle=\langle j^\phi (x) \rangle_{s} + \langle j^\phi (x) \rangle_{c} \  .
\end{eqnarray}
The first term, $\langle j^\phi (x) \rangle_{s}$, corresponds to the azimuthal current density in
the geometry of a straight cosmic having a magnetic flux running along its core,
and the second, $\langle j^\phi (x) \rangle_{c}$, is induced by the magnetic tube of radius $a$.

At this point we would like to analyze separately both contributions.

\subsubsection{Azimuthal current induced by a zero-thickness magnetic flux}

The azimuthal current induced by a magnetic flux running along the idealized cosmic string is given by,
\begin{eqnarray}
\label{Azimu}
\langle j^\phi (x) \rangle_{s}=-\frac{eq}{2\pi^2 r}
\int^{\infty}_{-\infty} dk \int^{\infty}_{0}\frac{d\lambda \lambda^2}
{\sqrt{\lambda^2 + k^2 + m^2}}\sum_{j}\epsilon_j J_{\beta_j}(\lambda r)J_{\beta_j+\epsilon_j}(\lambda r) \  .
\end{eqnarray}

The explicit calculation of this contribution was given in \cite{Mello13}. Here we briefly 
review its more important results. Using the identity bellow, 
\begin{equation}
\frac{1}{\sqrt{m^{2}+k^{2}+\lambda ^{2}}}=\frac{2}{\sqrt{\pi }}%
\int_{0}^{\infty }dt\ e^{-(m^{2}+k^{2}+\lambda ^{2})t^{2}} \  ,  \label{ident1}
\end{equation}
into \eqref{Azimu}, it is possible to integrate over the variable $\lambda$
by using the results form \cite{Grad}:
\begin{equation}
\int_{0}^{\infty }d\lambda \lambda ^{2}e^{-\lambda ^{2}t^{2}}J_{\beta
_{j}}(\lambda r)J_{\beta _{j}+\epsilon _{j}}(\lambda r)=\frac{%
e^{-r^{2}/(2t^{2})}}{4t^{4}}r\epsilon _{j}\left[ I_{\beta
_{j}}(r^{2}/(2t^{2}))-I_{\beta _{j}+\epsilon _{j}}(r^{2}/(2t^{2}))\right] \ .
\label{Int1}
\end{equation}
Introducing a new variable $y=r^{2}/(2t^{2})$, we get
\begin{equation}
\langle j^{\phi }\rangle _{s}=-\frac{eq}{2\pi ^{2}r^{4}}\int_{0}^{\infty }\
dy\ y\ e^{-y-m^{2}r^{2}/(2y)}\ [\mathcal{I}(q,\alpha _{0},y)-\mathcal{I}%
(q,-\alpha _{0},y)]\ ,  \label{j-cs1}
\end{equation}
where $\mathcal{I}(q,\alpha _{0},y)$ is defined by
\begin{equation}
\mathcal{I}(q,\alpha _{0},z)=\sum_{j}I_{\beta _{j}}(z)=\sum_{n=0}^{\infty }%
\left[ I_{q(n+\alpha _{0}+1/2)-1/2}(z)+I_{q(n-\alpha _{0}+1/2)+1/2}(z)\right]
,  \label{seriesI0}
\end{equation}%
and
\begin{equation}
\sum_{j}I_{\beta _{j}+\epsilon _{j}}(z)=\mathcal{I}(q,-\alpha _{0},z) \ .
\label{seriesI2}
\end{equation}
In the above development, we have used used the notation
\begin{eqnarray}
\label{alpha}
\alpha=eA_\phi/q =-\Phi/\Phi_0= n_0+\alpha_0 \  ,
\end{eqnarray}
with $n_0$ being an integer number. So we conclude that \eqref{j-cs1}
is an odd function of $\alpha_0$.

In \cite{Mello10} we have presented an integral representation for $\mathcal{I}(q,\alpha _{0},y)$:
\begin{eqnarray}
&&\mathcal{I}(q,\alpha _{0},z)=\frac{e^{z}}{q}-\frac{1}{\pi }%
\int_{0}^{\infty }dy\frac{e^{-z\cosh y}f(q,\alpha _{0},y)}{\cosh (qy)-\cos
(q\pi )}  \notag \\
&&\qquad +\frac{2}{q}\sum_{k=1}^{p}(-1)^{k}\cos [2\pi k(\alpha
_{0}-1/2q)]e^{z\cos (2\pi k/q)},  \label{seriesI3}
\end{eqnarray}
with $2p<q<2p+2$ and with the notation%
\begin{eqnarray}
f(q,\alpha _{0},y) &=&\cos \left[ q\pi \left( 1/2-\alpha _{0}\right) \right]
\cosh \left[ \left( q\alpha _{0}+q/2-1/2\right) y\right]  \notag \\
&&-\cos \left[ q\pi \left( 1/2+\alpha _{0}\right) \right] \cosh \left[
\left( q\alpha _{0}-q/2-1/2\right) y\right] \ .  \label{fqualf}
\end{eqnarray}
For $1\leqslant q<2$, the last term on the right-hand side of Eq. (\ref{seriesI3}%
) is absent.

By using the result (\ref{seriesI3}), and after the integration over $y$, the
expression (\ref{j-cs1}) is presented in the form
\begin{eqnarray}
\langle j^{\phi }\rangle _{s} &=&-\frac{em^{2}}{\pi ^{2}r^{2}}\ \left[
\sum_{k=1}^{p}\frac{(-1)^{k}}{\sin (\pi k/q)}\sin (2\pi k\alpha
_{0})K_{2}(2mr\sin (\pi k/q))\right.   \notag \\
&&\left. +\frac{q}{\pi }\int_{0}^{\infty }dy\frac{g(q,\alpha
_{0},2y)K_{2}(2mr\cosh y)}{[\cosh (2qy)-\cos (q\pi )]\cosh y}\right] \ .
\label{jazimu}
\end{eqnarray}
In the above expression $K_{\nu }(x)$ is the Macdonald function, and
\begin{eqnarray}
g(q,\alpha _{0},y) &=&\cos \left[ q\pi \left( 1/2+\alpha _{0}\right) \right]
\cosh \left[ q\left( 1/2-\alpha _{0}\right) y\right]   \notag \\
&&-\cos \left[ q\pi \left( 1/2-\alpha _{0}\right) \right] \cosh \left[
q\left( 1/2+\alpha _{0}\right) y\right] .  \label{gxy}
\end{eqnarray}

As we can see $\langle j^{\phi }\rangle _{s}$ depends only on the fractional part of the ration 
of the total flux by the quantum one, $\alpha_0$, and vanishes when this parameter is zero. We can 
say that this current is a manifestation of the Aharonov-Bohm effect.

\subsubsection{Core-induced azimuthal current}

The core-induced azimuthal current, $\langle j^\phi (x) \rangle_{c}$, can be written in a compact for by,
\begin{eqnarray}
\label{Azimu1}
\langle j^\phi (x) \rangle_{c}&=&\frac{eq}{(2\pi)^2 r}\int^{\infty}_{-\infty} dk
\int^{\infty}_{0}\frac{d\lambda \lambda^2 }{\sqrt{\lambda^2 + k^2 + m^2}} \nonumber \\
&&\times\sum_{j}\epsilon_j \tilde{J}_{\beta_j}(\lambda a)\sum^{2}_{l=1}\frac{H^{(l)}_{\beta_j}
(\lambda r)H^{(l)}_{\beta_j+\epsilon_j}(\lambda r)}{\tilde{H}^{(l)}_{\beta_j}(\lambda a)}  \   ,
\end{eqnarray}
where $H_\nu^l(x)$ with $l=1, 2$ represents the Hankel functions.  In order to develop this calculation, 
we rotate the integrals contour in the complex plane $\lambda$ as follows: by the angle $\pi/2$ for 
$l=1$ and $-\pi/2$ for $l=2$. By using the property below,
\begin{equation}
\label{Rel-V}
{\cal{V}}^{(i)}_j(\pm i\lambda, a)=\pm i{\rm Im}\{{\cal{V}}^{(i)}_j( i\lambda, a)\}  \  ,
\end{equation}
one can see that the integral over the segments $(0, \ i\sqrt{m^2+k^2})$ and $(0, \ -i\sqrt{m^2+k^2})$ 
are canceled. In the remaining integral over the imaginary axis we introduce the modified Bessel functions. 
Moreover, writing imaginary integral variable by $\lambda=\pm iz$, the core-induced azimuthal current reads,
\begin{eqnarray}
\label{Azimu2}
\langle j^\phi (x)\rangle_{c}&=&-\frac{eq}{\pi^3 r}\int_0^\infty dk \int^{\infty}_{\sqrt{k^2 + m^2}}
\frac{dz z^2}{\sqrt{z^2 - k^2 - m^2}} \nonumber \\
 &&\sum_{j}K_{\beta_j}(z r)K_{\beta_j + \epsilon_j}(z r)F^{(i)}_{j}(z a) \  ,
\end{eqnarray}
where we use the notation
\begin{eqnarray}
\label{F-ratio}
F^{(i)}_{j}(y)= \frac{I_{\beta_j+\epsilon_j}(y)-
\mbox{Im}[{\cal{V}}_j^{(i)}(iy/a, a)]I_{\beta_j}(y)}{K_{\beta_j+\epsilon_j}(y)+
\mbox{Im}[{\cal{V}}_j^{(i)}(iy/a, a)]K_{\beta_j}(y)} \  .
\end{eqnarray}
After a convenient coordinate transformations we write \eqref{Azimu2} as follow:
\begin{eqnarray}
\label{Azimu3}
\langle j^\phi (x)\rangle_{c}=-\frac{eq}{\pi^2 r^4}\int^{\infty}_{mr}z^2 dz \sum_{j} K_{\beta_j}(z)
K_{\beta_j + \epsilon_j}(z)F^{(i)}_{j}(z (a/r)) \  .
\end{eqnarray}

Before to start explicit numerical analysis related to the core-induced azimuthal current, 
let us now evaluate its behavior at large distance from the core. First we consider massive 
fields and in the limit $mr>>ma$.  In order to develop this analysis, we assume that the product 
$K_{\beta_j}(z)K_{\beta_j + \epsilon_j}(z)$ can be expressed in terms of 
their corresponding asymptotic forms. So we have the induced current density given below
\begin{equation}
\label{massive2}
\langle j^\phi\rangle_c \approx -\frac{eq}{2\pi r^4}\int^{\infty}_{mr} 
dz \ z e^{-2z}\sum_{j}F^{(i)}_{j}(z (a/r)) \  .
\end{equation}
The dominant contribution for this integral is given from the region near the lower limit of 
integration. Then the leading order contribution is,
\begin{equation}
\label{massive4}
\langle j^\phi(r)\rangle_c \approx -\frac{eqm^4}{4\pi (mr)^3}e^{-2mr}\sum_{j}F^{(i)}_{j}(m a) \  .
\end{equation}
The behavior of the zero-thickness azimuthal current, $\left\langle j_{\phi }\right\rangle _{c}$, 
decays as $e^{-2mr\sin(\pi/q)}/(mr)^{5/2}$ for $q>2$; consequently the latter dominates the  total azimuthal current at large distances.

Our next analysis is to study the behavior of azimuthal current for massless fields for $r>>a$. 
In order to do that, we shall use explicitly the radial functions in the region 
inside the tube. In  \cite{Souto1,Souto2} we provided exact solutions for $R_1(r)$ and $R_2(r)$ 
for the three different configurations of magnetic field. They are:
\begin{enumerate}
\item For the cylindrical shell:
\begin{eqnarray}
\label{30.01}
R^{(1)}_1(r)&=& J_{\nu_j}(\lambda r)\nonumber\\
R^{(1)}_2(r)&=&\hat{\epsilon}_j J_{\nu_j+\hat{\epsilon}_j}(\lambda r) \ ,
\end{eqnarray}
 where $\nu_j=q|j|-\frac{\tilde{\epsilon}_j}{2}$, with $\tilde{\epsilon}_j=1$ for $j>0$ and
$\tilde{\epsilon}_j=-1$ for $j<0$.

\item For the magnetic field proportional to $1/r$:
\begin{eqnarray}
\label{30.02}
R^{(2)}_1(r)&=&\frac{M_{\kappa,\nu_j}(\xi r)}{\sqrt{r}}\nonumber\\
R^{(2)}_2(r)&=&C^{(2)}_j \frac{M_{\kappa,\nu_j+\hat{\epsilon}_j}(\xi r)}{\sqrt{r}} \  ,
\end{eqnarray}
where
\begin{eqnarray}
\label{2.300}
\xi=\frac{2}{a}\sqrt{q^2\alpha^2 -\lambda^2 a^2} \  , \ \kappa=-\frac{2q^2 j \alpha}{\xi a}
\end{eqnarray}
and
\begin{equation}
\label{2.30.0}
C^{(2)}_j=\left\{\begin{array}{cc}
\frac{\lambda}{\xi}\frac{1}{(2q|j|+1)} ,  \  \ j>0  \ .
\\ -\frac{\xi}{\lambda}(2q|j|+1), \ j<0 \ . \end{array} \right.
\end{equation}
\item For the homogeneous magnetic field:
\begin{eqnarray}
\label{30.03}
R^{(3)}_1(r)&=&\frac{M_{\kappa-\frac{1}{4},\frac{\nu_j}{2}}(\tau r^2)}r \nonumber\\
R^{(3)}_2(r)&=&C^{(3)}_j \frac{M_{\kappa+\frac{1}{4},\frac{\nu_j+\hat{\epsilon}_j}{2}}(\tau r^2)}r \ ,
\end{eqnarray}
where $\tau$ and $\kappa$ are given by
where $\tau$ and $\kappa$ are given by
\begin{equation}
\label{tau}
\tau=q\alpha/a^2   , \  \ \ \kappa=\frac{\lambda^2}{4\tau}-\frac{qj}{2} \ ,
\end{equation}
with
\begin{equation}
\label{2.30.1}
C^{(3)}_j=\left\{\begin{array}{ccc} \frac{\lambda}{\sqrt{\tau}}\frac{1}{2q|j|+1} \
\ j> 0 \ .
\\ -\frac{\sqrt{\tau}}{\lambda}(2q|j|+1), \ j<0 \ . \end{array}\right. \
\end{equation}
\end{enumerate}

For the second and third models, the radial functions are given in terms of
Whittaker functions, $M_{\kappa,\nu}(z)$ \cite{Grad}.

Now we would like to discuss the behavior of the core-induced azimuthal current at 
large distance of the core, considering massless fields. Instead to use the summation 
on the angular moment $j$ in \eqref{Azimu3}, we use $n=j-1/2$. In this way, we shall use 
a new notation, $F^{(i)}_{n} \equiv F^{(i)}_{j}$. Moreover, we change $n$ by $n-n_0$, 
being $n_0$ given in \eqref{alpha}. So from \eqref{Azimu3} we can write,
\begin{eqnarray}
\label{Azimu4}
\langle j^\phi (x)\rangle_{c}=-\frac{eq}{\pi^2 r^4}\int^{\infty}_{0} d z z^2 
\sum_{n=-\infty}^\infty K_{\beta}(z)
K_{{\tilde{\beta}}}(z)F^{(i)}_{n-n_0}(z (a/r))  \ .
\end{eqnarray}
In the above expression we are using the notation:  
\begin{eqnarray}
\label{F-ratio1}
F^{(i)}_{n-n_0}(z (a/r))= \frac{I_{{\tilde{\beta}}}(z (a/r))-
\mbox{Im}[{\cal{V}}_{n-n_0}^{(i)}(iz, (a/r))]I_{\beta}(z (a/r))}{K_{{\tilde{\beta}}}(z (a/r))+
\mbox{Im}[{\cal{V}}_{n-n_0}^{(i)}(iz, (a/r))]K_{\beta}(z (a/r))} \  .
\end{eqnarray}
In \eqref{F-ratio1} the orders of Bessel functions are given by
\begin{eqnarray}
\label{beta}
\beta=q|n+1/2+\alpha_0|-\frac12\frac{|n+1/2+n_0|}{n+1/2+\alpha_0} \ , \nonumber\\
{\tilde{\beta}}=q|n+1/2+\alpha_0|+\frac12\frac{|n+1/2+n_0|}{n+1/2+\alpha_0} \ .
\end{eqnarray}

Expanding the integrand of \eqref{Azimu4} in powers of $a/r$, we keep only the dominant term 
that is given by the smaller power of this ratio. For this analysis we have two possibilities: 
for $\alpha_0>0$ ($0\leq\alpha_0<1/2$) this term is given by $n=-1$, and for 
$\alpha_0<0$ ($-1/2<\alpha_0\leq 0$) this term is given for $n=0$. 

Moreover, using the expansions for the modified Bessel functions for small arguments \cite{Abramo}, the leader contributions are:
\begin{itemize}
\item For $\alpha_0>0$:
\begin{equation}
\label{b6}
F^{(i)}_{-1-n_0}\left(z\frac{a}{r}\right)\approx \frac{2}{\Gamma^2(\beta)}
\frac{1+i{\cal{V}}_{-1-n_0}^{(i)}(iz, a/r)\left(z\frac{a}{r}\right)
\left(\frac{az}{2r\beta}\right)}{\frac{\Gamma(1-\beta)}
{\Gamma(\beta)}-i{\cal{V}}_{-1-n_0}^{(i)}(iz, a/r)
\left(z\frac{R}{r}\right)\left(\frac{az}{2r}\right)\left(\frac{2r}{az}\right)^{2\beta}} \  .
\end{equation}
\item For $\alpha_{0}<0$, and
\begin{equation}
\label{b7}
F^{(i)}_{-n_0}\left(z\frac{a}{r}\right)\approx -\frac{2}{\beta\Gamma^2(\beta)}
\frac{1+i{\cal{V}}_{-n_0}^{(i)}(iz, a/r)\left(z\frac{R}{r}\right)\left(\frac{2r\beta}{az}\right)}
{\left(\frac{2r}{az}\right)^{2\beta}-i\frac{\Gamma(1-\beta)}
{\Gamma(\beta)}{\cal{V}}_{-n_0}^{(i)}(iz, a/r)\left(z\frac{R}{r}\right)\left(\frac{az}{2r}\right)}  \  .
\end{equation}
\end{itemize}

The next steps are the calculations of the dominants contribution for
the coefficient that contains all the information about the core, ${\cal{V}}_{-1-n_0}^{(i)}(iz, a/r)$ 
and ${\cal{V}}_{-n_0}^{(i)}(iz, a/r)$, for the three models. This can be done by explicit 
substitution of the radial functions, $R^{(i)}_1(iz, a/r)$ and $R^{(i)}_2(iz, a/r)$,
into \eqref{Z.Bessel}.  So, after some intermediate steps, we find:
\begin{equation}
\label{b13}
\langle j^\phi (r) \rangle_{c}\approx 2\frac{|\alpha_0|}{\alpha_0}\frac{eq}{\pi^2 r^4}\frac{\beta-\chi^{(l)}}
{\left(\frac{2r}{a}\right)^{2\beta}\chi^{(l)}}\frac{\beta}{2\beta+1} \  ,
\end{equation}
where
\begin{equation}
\label{b4-a}
\beta=q\left(\frac{1}{2}-|\alpha_{0}|\right)+\frac{1}{2}  \
\end{equation}
and $\chi^{(l)}$ is a parameter depending on the specific model adopted for the
magnetic field, given bellow by:
\begin{equation}
\label{b12}
\chi^{(l)}=\left\{\begin{array}{ccc}
\nu=q|n_0-\frac12\frac{|\alpha_0|}{\alpha_0}|-\frac12\frac{|\alpha_0|}{\alpha_0}, \ \mbox{for the model ({\it{i}})}\\
q\alpha(q+1)\frac{M_{\frac{q}{2}\frac{|\alpha_0|}{\alpha_0},\nu}(2q\alpha)}
{M_{\frac{q}{2}\frac{|\alpha_0|}{\alpha_0},\nu+1}(2q\alpha)}, \ \mbox{for the model ({\it{ii}})}\\
\frac{\sqrt{q\alpha}}{2}(q+1)\frac{M_{-\frac{1}{2}\frac{|\alpha_0|}{\alpha_0}
\frac{q+1}{2},\frac{\nu}{2}}(\tau R^2)}{M_{-\frac{1}{2}\frac{|\alpha_0|}
{\alpha_0}\frac{q+1}{2},\frac{\nu}{2}}(\tau R^2)}, \ \mbox{for the model ({\it{iii}})} \ .
\end{array}\right.
\end{equation}	
On basis of the above results we can say that for the three models considered, 
the core-induced azimuthal current density decays with, $\frac1{r^4(r/a)^{2\beta}}$,
for large distance from the tube. In \cite{Mello13} we have shown  that for massless fields  
the zero-thickness azimuthal current decays with $1/r^4$. So, we conclude that for large distance 
from the core, the total azimuthal current, \eqref{Azimuthal3}, is dominated by the zero-thickness  contribution.

The next investigation is the behavior of $\langle j^\phi (r) \rangle_{c}$ near the core, for the 
three models. In general the current diverges in this region. The dominant. To find the leading 
term it is convenient to introduce a new variable $z=\beta_j x$, and 
use the uniform expansion for large order for the modified Bessel functions \cite{Abramo}.
However, before to do that, we would like to notice that, 
changing $n\rightarrow -n-1$ the summation over \textit{j} keeps 
unchanged, but the parameter $\nu_j$ change as 
$\nu_j\rightarrow\tilde{\nu}_j$ and $\tilde{\nu}_j\rightarrow\nu_j$. If, in addition,
we also change $\alpha\rightarrow -\alpha$, then $\beta_j\rightarrow\tilde{\beta}_j$ 
and $\tilde{\beta}_j\rightarrow\beta_j$.  It means that, when we make $n\rightarrow -n-1$ 
and $\alpha\rightarrow -\alpha$ we have $ F^{(i)}_{j}(y)\longrightarrow- F^{(i)}_{j}(y)$.
On basis of this analysis, and considering $\alpha>0$, the behavior of the
core-induced azimuthal current near the boundary is given by, 
\begin{eqnarray}
\label{c4}
\langle j^\phi (x) \rangle_{c}\approx 2\frac{eq}{\pi^2 r^4}\sum_{n>0}\beta^{3}_{j}
\int^{\infty}_{\frac{mr}{\beta_j}}dx \ x^2 F^{(i)}_{j}(\beta_j x({a}/{r}))
K_{\beta_j}(\beta_j x)K_{\beta_j + \epsilon_j}(\beta_j x) \ .
\end{eqnarray}
Because we are considering $n>>1$, from now on we use the approximation, $\beta_j\approx\nu_j\approx qn$.

For the three models, it is necessary to find the leading term of $F^{(i)}_j$ for
large value of $j$. We have shown in \cite{Souto1}, that for the three models the leading terms have the same structure given below,
\begin{eqnarray}
\label{c6}
F^{(1,2)}_{j}(qn x(a/r))\approx \frac{1}{4q^2\pi}\frac{e^{2qn\tilde{\eta}}}{ n^2 (1+e^{2\tilde{\eta}})} \  ,
\end{eqnarray}
where $\tilde{\eta}=\sqrt{1+x^2(a/r)^2}$. 

Finally for the  three models, the core-induced azimuthal current density is given by
\begin{eqnarray}
\label{c7}
\langle j^\phi(x) \rangle_c \approx \frac{eq}{4\pi^2 r^4}\sum_{n>0}\int^{\infty}_{\frac{mr}{qn}}
dz \ z^2\frac{e^{-2qn(\eta-\tilde{\eta})}}{(1+e^{2\tilde{\eta}})\sqrt{1+z^2}} \ ,
\end{eqnarray}
where $\eta=\sqrt{1+x^2}$. Using the approximation $\eta-\tilde{\eta}\approx z(1-a/r)$, and 
observing that the denominator of the integrand in \eqref{c7} can be approximate to unity, for the 
three models, we have:
\begin{eqnarray}
\label{c8}
\langle j^\phi(x)\rangle_c \approx \frac{eq}{4\pi^2 r^4}\sum_{n>0} \int^{\infty}_{\frac{mr}{qn}}
dz \ z^2 e^{-2qn(1-a/r)}\ .
\end{eqnarray}
Solving the above integral we have
\begin{eqnarray}
\label{c9}
\langle j^\phi(r) \rangle_c \approx \frac{e}{(4\pi q )^2}\frac1 r\frac1{(r-a)^3} \  . 
\end{eqnarray}
Here, we must notice that the current density diverges near the boundary. 

Because the zero-thickness azimuthal current, $\langle j^\phi(r) \rangle_s$, presents a finite value
near the boundary, we can conclude that, in this region, the total azimuthal current, 
\eqref{Azimuthal3} is dominated by the core-induced contribution. 

\begin{figure}[!htb]
\begin{center}
\includegraphics[width=0.4\textwidth]{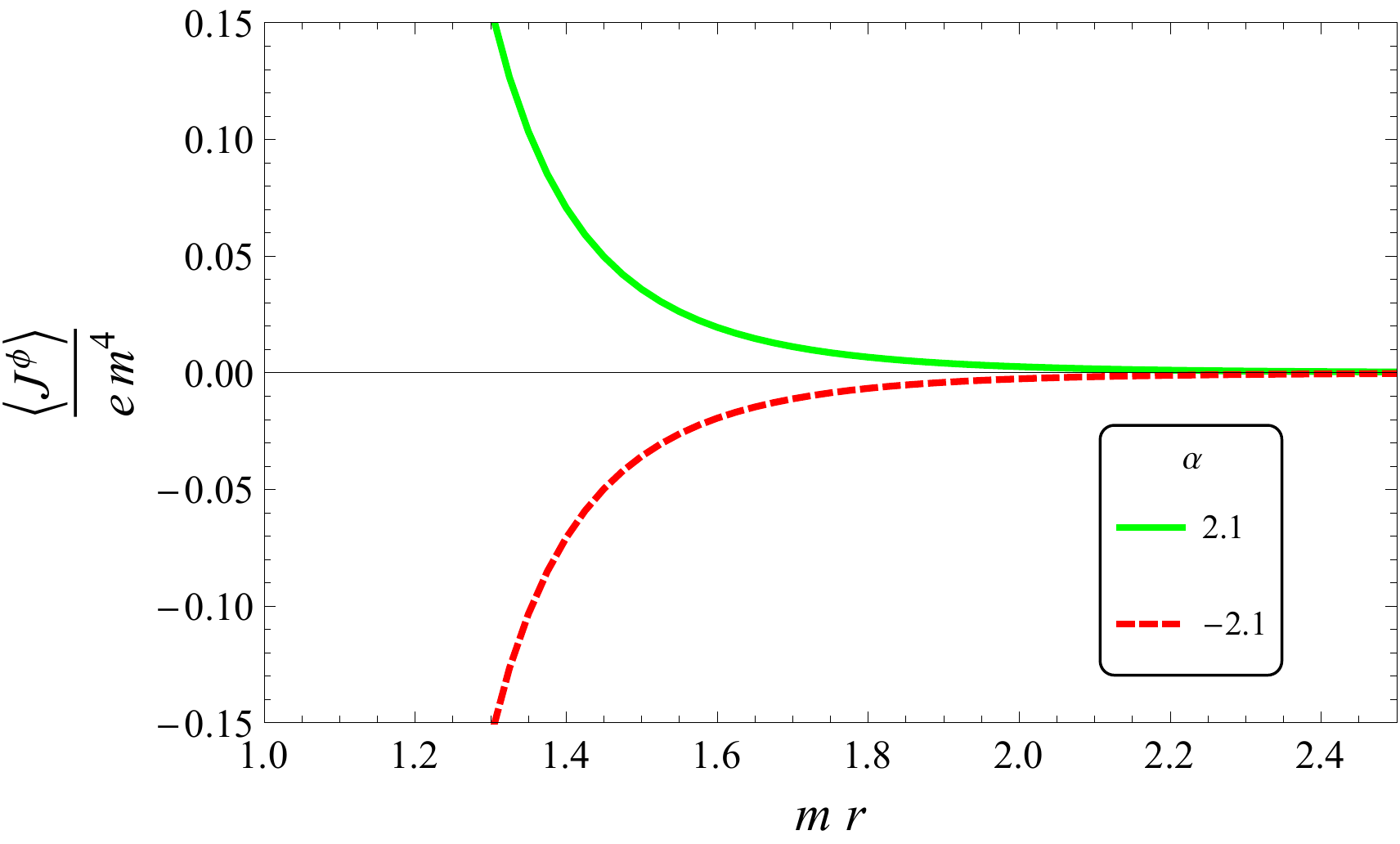}
\includegraphics[width=0.4\textwidth]{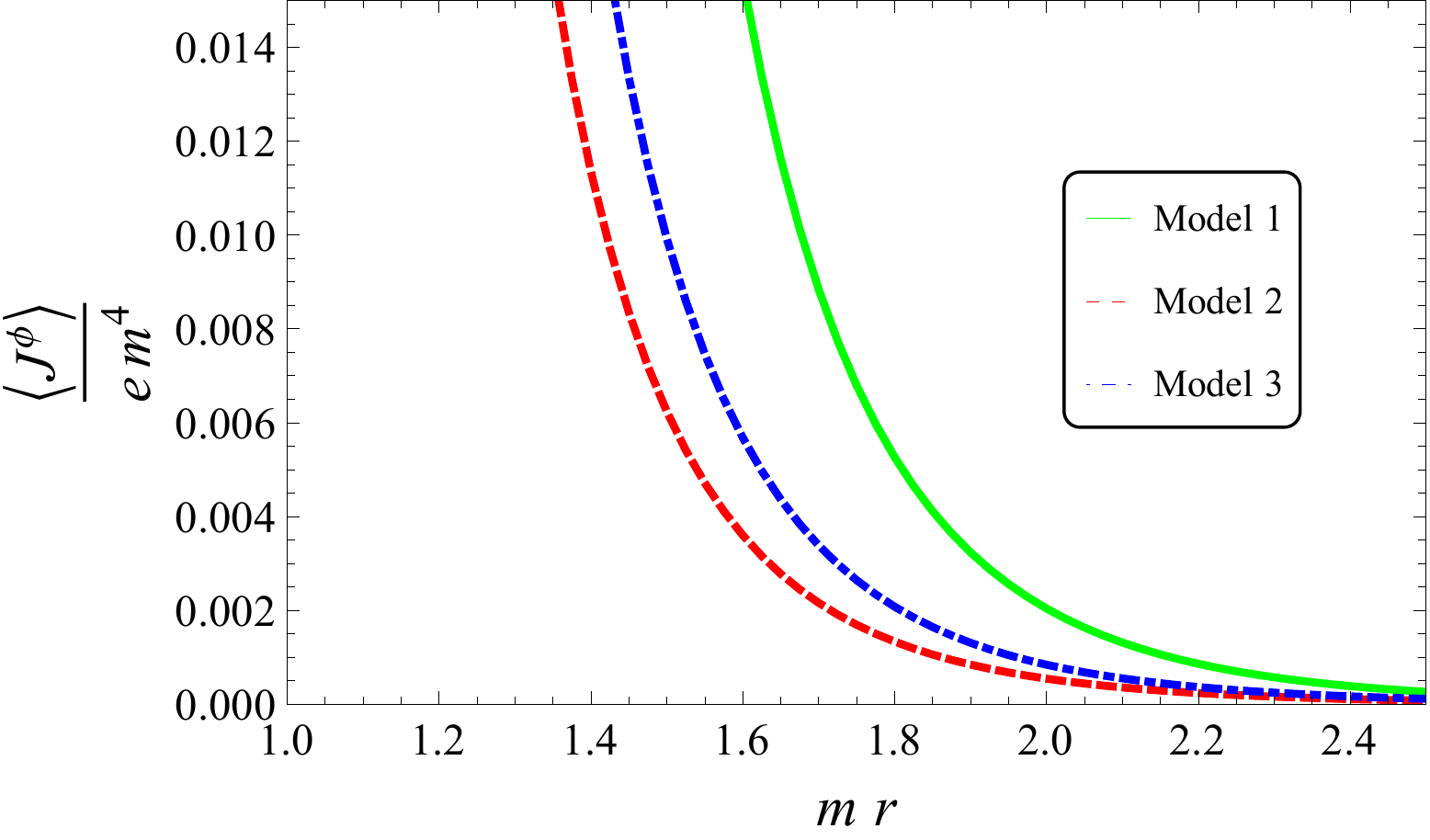}
\caption{The core-induced azimuthal current density is plotted, in units of $m^4e$ as function of 
	$mr$ for $q=1.5$ and $ma=1$. On the left plot we consider the current induced by the magnetic 
	field configuration of the first model, and taking $\alpha=2.1$ and $\alpha=-2.1$. On the right 
	plot we display the intensity of the core-induced current for the three different models of 
	magnetic fields considering $\alpha=2.1$.}
\label{fig1}
\end{center}
\end{figure}

Let us now present some additional informations which are not provided  by the analytical expressions. 
In Fig. \ref{fig1} we exhibit the dependence of the core-induced azimuthal current 
density with $mr$ considering $q=1.5$ and $ma=1$. On the left plot, we present the behavior for 
the current induced by the cylindrical shell of magnetic field, taking into account positive and 
negative value for $\alpha$. On the right plot we exhibit its behavior as function of $mr$,  for 
the three different models of magnetic fields considering  $\alpha=2.1$. By this plot we can infer 
that for a given  point outside the tube, the intensity of the current induced by the first model is  the biggest one. 

\begin{figure}[!htb]
	\begin{center}
		\includegraphics[width=0.4\textwidth]{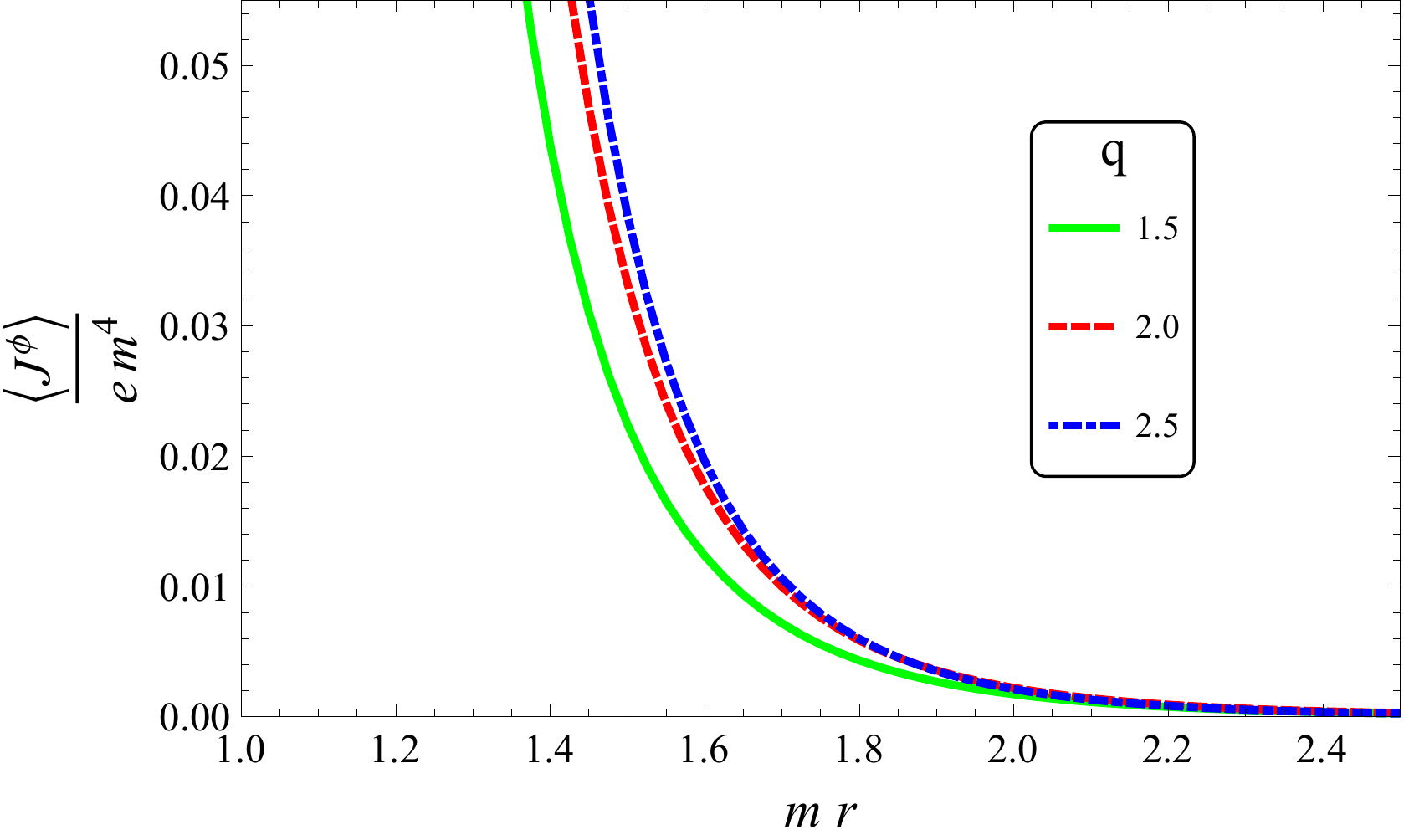}.
		\caption{The core-induced azimuthal current density in units of $m^4e$, 
			due to a magnetic field concentrated on a shell, as function of $mr$ 
			considering $q=1.5, \ 2.0, \ 2.5$. In this plot we adopted $\alpha=1.2$.}
		\label{fig2}
	\end{center}
\end{figure}

In Fig \ref{fig2}, we exhibit, for the magnetic field concentrated in a cylindrical shell, the behavior of $\langle j^\phi(r) \rangle_c$ as function of $mr$ considering 
$q=1.5, \ 2.0, \ 2.5$. By this plot we can infer that the intensity of 
the current increases with $q$. Here we adopted $\alpha=1.2$ and $ma=1$.

\begin{figure}[!htb]
\begin{center}
\includegraphics[width=0.4\textwidth]{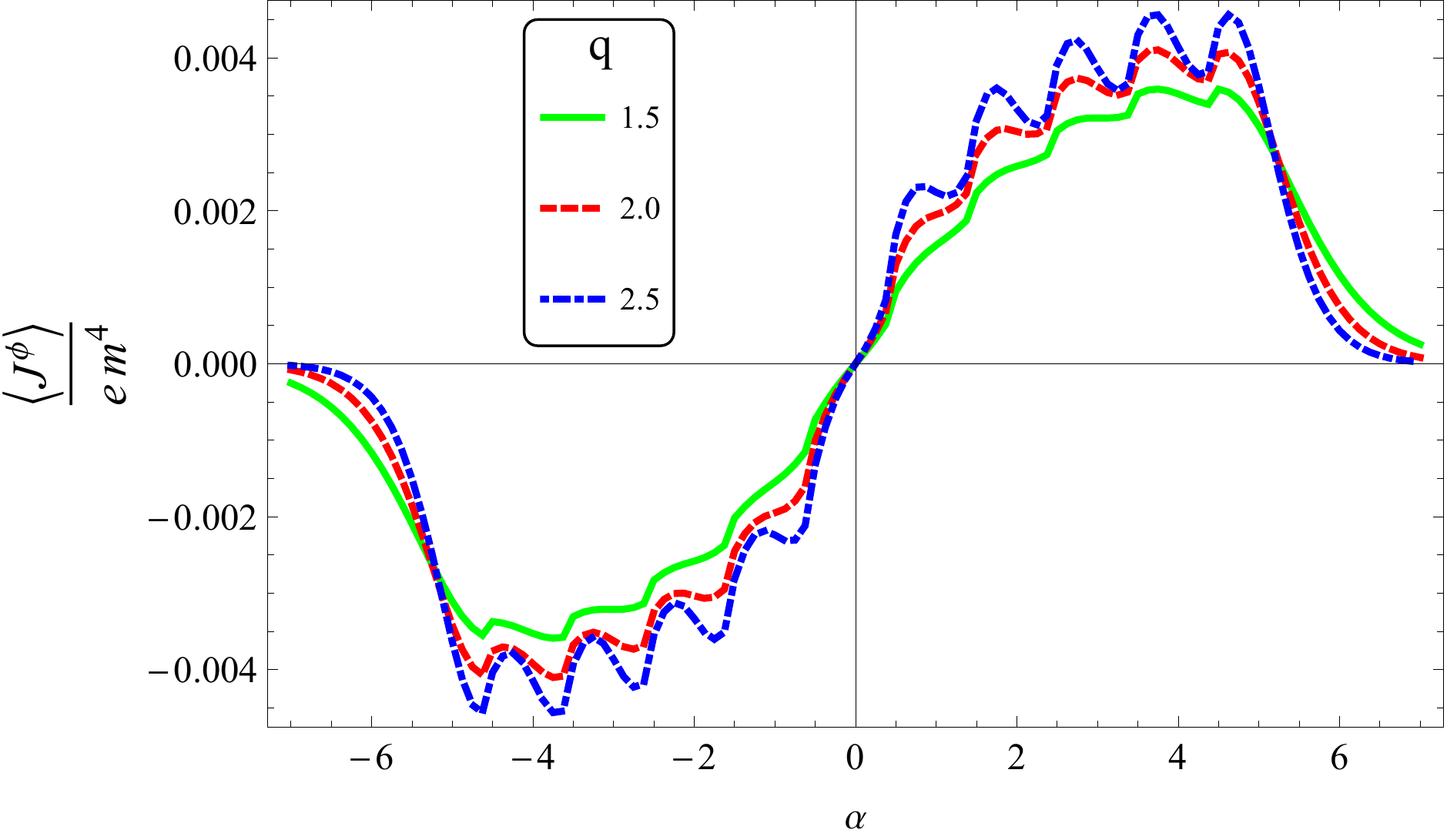}
\includegraphics[width=0.4\textwidth]{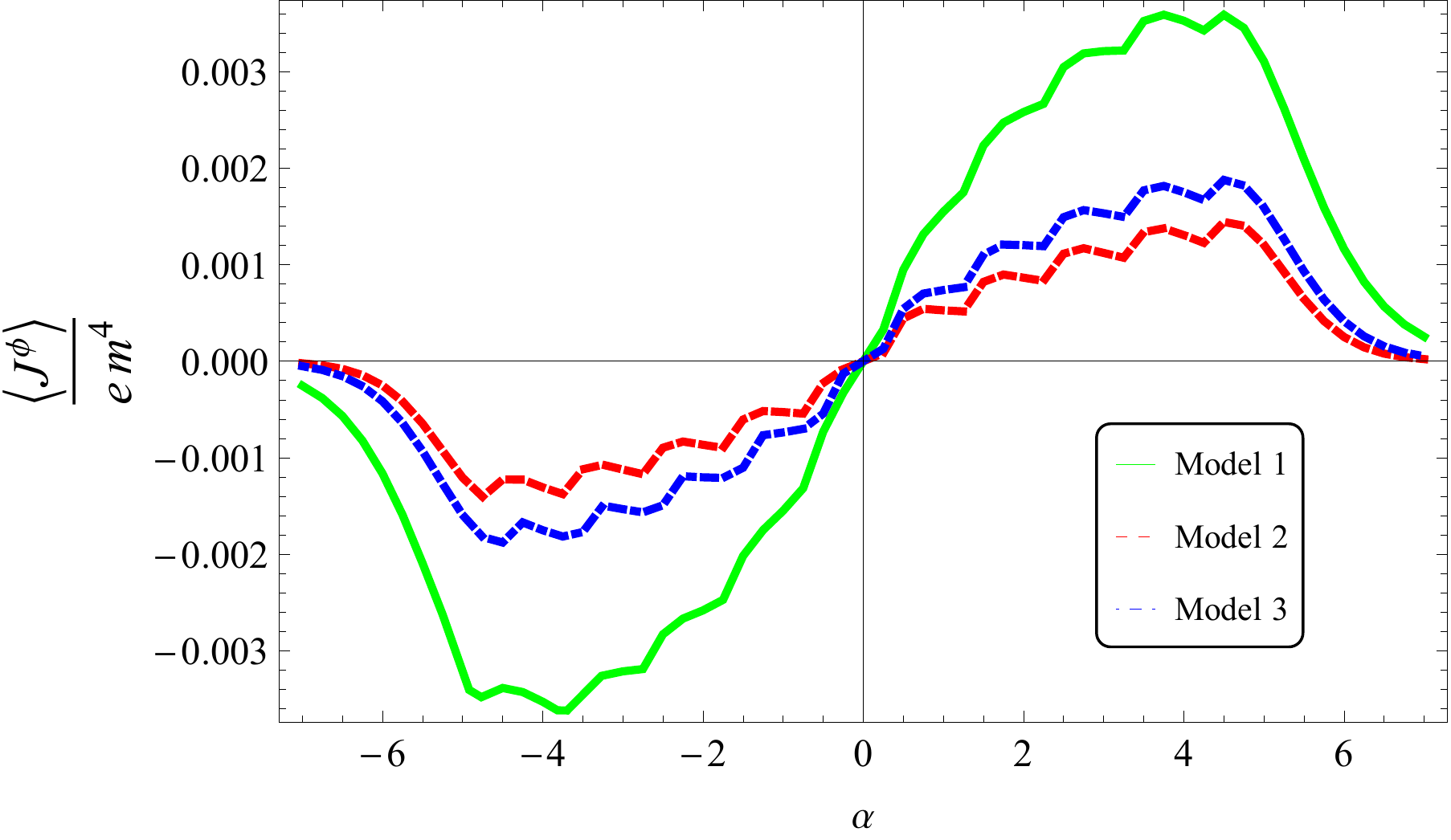}
\caption{The core-induced azimuthal current density is plotted in units of $m^4e$, as function of 
	$\alpha$ for $ma=1$ and $mr=2$. On the left plot, we exhibit only the current induced by the 
	first model considering different values of $q$, and on the 
right plot, we exhibit the current induced by the three different models  of magnetic fields, considering $q=1.5$.}
\label{fig3}
\end{center}
\end{figure}

In Fig \ref{fig3} we exhibit the behavior of the core-induced current
as function of $\alpha$, considering $ma=1$ and $mr=2$. On the left plot we display 
the current induced by the first model of magnetic field for $q=1.5, \ 2.0, \ 3.5$.
On the right plot we consider the currents induced by the three different models,
adopting $q=1.5$. For both plots, we assume that $\alpha$ varies in the interval $[-7.0, \ 7.0]$.  By them we can infer once more that, the intensity of the current increases with 
$q$ (left plot) and the first model provide the current with biggest intensity (right plot).

\section{Conclusions}
\label{Conclu}

In this paper we have analyzed the influence of the conical topology of the spacetime 
and the finite core effect, on the vacuum expectation value of fermionic azimuthal current,
induced by magnetic fluxes of finite extension. In the model considered we have adopted that
the geometry of the spacetime corresponds to an idealized cosmic string everywhere, surrounded 
by a magnetic tube of radius $a$. In this tube three different configurations of
magnetic fields have been taken into account: a cylindrical shell, a field decaying 
with $1/r$ and finally a homogeneous magnetic field. In order to develop these 
analysis we had to construct the normalized fermionic wave-functions for the 
region outside the tube, and calculated the fermionic azimuthal current density 
by using the mode summation method. 
 
Although the magnetic field vanishes outside the tube, the magnetic field inside induces a 
non-vanishing azimuthal current density in the exterior region. This phenomenon is explicitly 
manifested in the structure of the induced current. The latter has been decomposed in two distinct 
contributions: The first one corresponds to the idealized cosmic string with a magnetic flux running along its core. It is given in \eqref{jazimu} and is a periodic function of the total flux with the period equal to quantum flux, $\Phi_0=2\pi/e$. 
The second contribution named core-induced given by \eqref{Azimu3}, takes into account a specific configuration for the magnetic field inside the tube, and  in general,
is not a periodic function of the magnetic flux and depends on the total magnetic 
flux inside the core. 

 By general analysis we could observe that the core-induced azimuthal current decays with $e^{-2mr}/(mr)^3$ for $r>>a$ (see \eqref{massive4}). Because the corresponding zero-thickness azimuthal current decays as $e^{-2mr\sin(\pi/q)}/(mr)^{5/2}$, the latter dominates the total azimuthal current. For massless field and at large distance from the core, the result was given by Eq. \eqref{b13}, and explicitly we see that this current decays with $\frac1{r^4(r/a)^{2\beta}}$. Comparing this result with the corresponding one for the zero-thickness azimuthal current
which decays with $1/r^4$, we conclude that for large distance the total azimuthal
current is dominated by the zero-thickness contribution.
Finally for point near the tube core, the core-induced current 
diverges with $\frac1{r^4(1-a/r)^3}$, as shown by \eqref{c9}. So this contribution
is dominant in this region. 

Finally we have also provide, by using numerical evaluation, the
behavior of the core-induced current as function of several 
physical quantities relevant in our analysis. In Fig. 1 we have two plots.
In the left plot an expected result is presented. The current changes its sign
when we change the sign of $\alpha$. In the right plot, it is exhibited
the behavior of the current for the three models of magnetic field as function of $mr$. It
is shown that the intensity of the current induced by the first model is the largest one.
In Fig. 2, we exhibit the behavior of current density for the
first model as function of $mr$, for fixed value of $\alpha$ and different
values of $q$. By this plot we can infer that the intensity of the current 
increasing with $q$. Finally Fig. 3, we have displayed  
current as function of the intensity of the magnetic flux. In the left
plot we have considered only the first model fixing $mr$ and 
varying $q$. This plot reinforces the fact that the intensity of the current
increases with $q$; moreover, it shows that the current is not a periodic 
function of the flux. The right plot exhibits the behavior of the
core-induced current, for the three models, for fixed value of $q$. 
Also this plot reinforce that the first model provides the current with 
biggest intensity.

\textbf{Acknowledgments}: M. S. M. S. thanks Coordenação de Aperfeiçoamento de
Pessoal de Nível Superior (CAPES) for financial support.
E. R. B. M. thanks Conselho Nacional de Desenvolvimento
Científico e Tecnológico (CNPq), Process No. $313137/
2014-5$, for partial financial support.

\end{document}